\documentstyle[]{article}
\textwidth=160mm \textheight=225mm \vspace{1cm}
\title{The Dark Energy and the Fate of Universe}
\author{Mian WANG \\
{\small Department of Physics, Henan Normal University} \\
{\small Xinxiang, Henan, China  453002}\\ {\small e-mail:
wangm@henannu.edu.cn }}
\date{}

\begin{document}
\maketitle
\def\be{\begin{equation}}
\def\ee{\end{equation}}
\def\vp{\vspace{5 mm}}
\def\f{\frac}
\vspace{2cm}
\begin{abstract}
Recent observations confirm that our universe  is flat
 and  consists of a dark energy component $\Omega_{DE}\simeq 0.7$.
 This  dark energy is responsible for the cosmic
acceleration as well as determines the feature of future evolution
of the universe. In this paper, we discuss the dark energy of
universe in the framework of scalar-tensor cosmology.  It is shown
that the dark energy is the main part of the energy density of the
gravitational scalar field
and the future universe will expand as $a(t)\sim t^{1.3}$.\\
\\
PACS number(s): 98.80.Cq, 04.50.+h
\end{abstract}
\vspace{3cm}

Observations of large scale structure, the Hubble diagram of type
Ia supernovae and the angular power spectrum of the cosmic
microwave background {\sl et al}, all indicate that the universe
is flat and is undergoing cosmic acceleration by virtue of  a dark
energy component, $\Omega_{DE}\simeq 0.7$, with negative pressure
$[1]$. It is supposed that the dark energy may consist of a
cosmological constant (vacuum energy) or the quintessence, such as
a scalar field with negative pressure $[2]$.  It is important to
investigate the nature of this energy component, because it is not
only responsible for the present accelerated expansion but also
will determinate the fate of our universe.  For example, if it is
a cosmological constant, the universe will expand exponentially in
the future.  On the other hand, if the dark energy is the energy
of a slowly changing scalar field  $\varphi$ with $w_\varphi\simeq
-1$ and the potential  $V(\varphi)$ driving the present stage of
acceleration decreases slowly and vanishes eventually, the speed
of expansion decreases after a transient de Sitter-like stage and
reaches Minkowski regime $[2]$,  or the potential falls to
$V(\varphi)=-\infty$, the universe eventually collapse, even if it
is flat$[3]$. In this letter we investigate the dark energy in the
framework of scalar-tensor cosmology$[4]$
with an added potential term. \\
\vp

{\sl The scalar-tensor cosmology}\ \ \ \ We should assume a
modified action of $[4]$ by adding a $V(\phi)$ term as follows
 \be
   {\cal A}=\int d^4x \sqrt{-g}\,\,\left[-\phi R
    - \frac{\omega}{\phi} \phi\cdot\phi -2 \phi \lambda(\phi)
   -2 \phi V(\phi)
      -\frac{\Gamma (u\cdot\phi)^2}{1-\phi}+16\pi L_m\right] ,
   \label{fb1}
 \ee
 in which $L_m$ is the Lagrangian of matter,
$\phi\cdot\phi\equiv\phi_{,\sigma} \phi^{,\sigma}$,
 $u\cdot\phi\equiv u_{\mu}\phi^{,\mu}$,
 $u_{\mu}$ is the four-velocity, $\phi^{,\mu}\equiv
\partial\phi/\partial x_{\mu}$, and $\Gamma$ is a constant of
 mass dimension 0. The $\Gamma-$term describes the coupling of the
 field $\phi$ with matter which is described as an ideal fluid.
The coupling function $\omega(\phi)$ and the cosmological function
$\lambda(\phi)$ are given as
 \be 2\omega(\phi)+3 =
\frac{\xi}{1-\phi}\, , \label{fb2} \ee \be \lambda(\phi) = 2\xi
\beta (1-\phi- \phi \,\, ln\,\phi) , \label{fb3}
 \ee
 where $\xi$ and $\beta$ are two dimensionless constants, here we
  set $\xi=7.5,\,\,\beta=1.7\times 10^{-16}$ as $[4]$ and
  $\Gamma=0.06$. The  potential
$V(\phi)$ is \be
        V(\phi)=\beta^2\sqrt{1-\phi} \; \; e^{-\alpha\phi}\label{fb4}
\ee in which $\alpha$ is a constant of order $O(10^2)$, it is set to be 81.5
in the model. \\

Then the Friedmann equations read \be
H^2=\frac{8\pi}{3\phi}(\rho_{m}+\rho_\phi)-\frac{k}{a^2} \, ,
\label{fb5}
 \ee
\be
 \frac{\ddot a}{a}=-\frac{4\pi}{3\phi}(\rho_{m}+\rho_\phi+ 3p_m+3p_\phi)\, ,
\label{fb6}
 \ee
and
\begin{eqnarray}
\ddot{\phi} + 3H \dot{\phi} &+& \frac{\dot{\phi}^2}{2 (1-\phi)}
 + \frac{8\pi(3p_m-\rho_m)}{2\omega+3}-\frac{\Gamma}{2\omega+3}
 \frac{\phi\dot\phi^2}{(1-\phi)^2}
 -\frac{2\Gamma}{2\omega+3}\frac{\phi\ddot\phi}{1-\phi} \nonumber \\
 &=&4\beta \phi (1-\phi^2)+\frac{2}{\xi}\phi\sqrt{1-\phi} \;\beta^2 \; e^{-\alpha\phi}
 \left(1-\frac{1}{2}\phi+\alpha(1-\phi)\right),
\label{fb7}
\end{eqnarray}
\be \dot{\rho}_m + 3 H(\rho_m + p_m)
=\frac{\Gamma}{4\pi}\,\left(\frac{\dot\phi\ddot\phi}{1-\phi}
+\frac{1}{2}\frac{\dot{\phi}^3}{(1-\phi)^2}+3H\frac{\dot\phi^2}{1-\phi}
\right) , \label{fb8}
\ee where $\rho_{m}$ , $p_m$ and $\rho_\phi$
, $p_\phi$ are energy density and pressure of the matter  and
field $\phi$
 respectively. The expressions are
  \be
\rho_{\phi}=\frac{1}{8\pi} \left(\phi \lambda + \phi V
  +\frac{\omega
  \dot{\phi}^2}{2\phi}-3 H \dot{\phi}
  -\frac{3}{2}\frac{\Gamma\dot{\phi}^2}{1-\phi}\right) , \label{fb9}
\ee
 \be p_{\phi} = \frac{1}{8\pi} \left(-\phi \lambda -\phi V+
\frac{\omega\dot{\phi}^2}{2\phi}+2 H \dot{\phi}+\ddot{\phi}
-\frac{1}{2}\frac{\Gamma\dot{\phi}^2}{1-\phi}\right) .
\label{fb10} \ee
\\
\vp

{\sl Exponential inflation} \ \ \ \  When the field $\phi$ rolls
down the potential hill from $\phi\simeq 0$, the universe expands
exponentially as described in $[4]$ since $V(\phi)$ can be
neglected when $\phi<1$. The Hubble parameter $H$ is determined by
the value of $\lambda(\phi)\simeq 2 \xi\beta$ when $\phi\ll 1$. In
the duration of inflation, the energy density of the scalar field
$\rho_\phi$ dominates and $p_\phi=-\rho_\phi$.
It is evident that $\rho_\phi$ acts as a cosmological constant.\\
\vp

{\sl Power-law expansion} \ \ \ \  As $\phi$ increases, the
$\dot\phi^2$ term in equation(9) increases also, therefore
$w_\phi\equiv p_\phi/\rho_\phi$ increases quickly. The inflation
ends dynamically when $\rho+3p\simeq \rho_\phi+ 3p_\phi$ increases
from $-2\rho_\phi$ to zero ($\phi\sim 1$). Then the universe
begins the era of expansion according to power law $a\sim
t^{2/3},$
 because the energy density $\rho_\phi$ remains to be dominating
 and the scalar $\phi$ behaves as massive particle with effective
 mass $\sim 2\sqrt{\beta}$ when $\phi$ goes near to the bottom of potential
 well. That is to say, the gravitational scalar field $\phi$
 behaves as vacuum energy in the era of inflation and as cold dark
 matter after inflation.\\

 After inflation, $\phi$ approaches near to its bottom value $1$. Let
$$
   \phi=1-\sigma^2,
$$
 it follows
\be
   (1-2\gamma)\ddot{\sigma}+3H\dot{\sigma}+4\beta\sigma=
    -sign(\sigma)\frac{\beta^2e^{-\alpha}}{2\xi},\label{fb11}
\ee
 in which
$$
        \gamma\equiv\frac{\Gamma}{\xi}.
$$
  From equation (11), we see that the $\sigma(t)$ oscillates with frequency
  $\omega_0$:
$$
           \omega_0^2=\frac{4\beta}{1-2\gamma}.
$$
Then we set
$$
           \sigma(t)=f(t)\; \cos\,\omega_0 t,
$$  we have equations
 \be
            \dot f =-\frac{3H}{2(1-2\gamma)}f, \label{fb12}
\ee\be
          \ddot f=-\frac{3H}{1-2\gamma}\dot
          f-\frac{2\beta^2e^{-\alpha}}{\pi\xi}, \label{fb13}
\ee and
 \be
      \dot{\rho}_m + 3 H(\rho_m + p_m)
      =\frac{\Gamma}{\pi}\,\left(\frac{3\beta^2e^{-\alpha}}{2\pi\xi(1-2\gamma)}
       H f+\frac{3\beta}{1-2\gamma}Hf^2 \right) . \label{fb14}
\ee
 The energy density of $\phi$ consists of $\rho_{DM}$ and
 $\rho_{DE}$:
\be
           \rho_{DM}= \frac{\xi\beta}{2\pi}(1-2\gamma)f^2,\ \
           \ \ \  \rho_{DE}=\frac{\beta^2e^{-\alpha}}{4\pi^2}f  \label{fb15}
\ee and the pressure of $\phi$ is
\be
           p_\phi=-(1-\frac{1}{\xi})\frac{2\beta^2e^{-\alpha}}{4\pi^2}f. \label{fb16}
\ee

 We notice that in equation(15), the multiplier constants
 $\frac{\xi\beta}{2\pi}(1-2\gamma)=O(10^{-16})$ and
 $\frac{\beta^2e^{-\alpha}}{4\pi^2}=O(10^{-69})$. Therefore, when
 $f>10^{-53}$, the $\rho_{DM}$ dominates, equation(12) has
 solution
 \be
               f=\frac{1}{\sqrt{3\xi\beta}t} \label{fb17}
\ee and
\be
          H=\frac{2(1-2\gamma)}{3t},\ \ \ \ \ \ a(t)\sim t^
          {\frac{2}{3}(1-2\gamma)}. \label{fb18}
\ee \\

{\sl Deceleration-acceleration transition } \ \ \ \   When $f(t)$
decreases to less than $10^{-53}$, correspondingly
$t>10^{60}M_P^{-1}$, the dark energy dominates gradually. In this
transient stage, we have not analytic solutions. From numerical
solutions by computer, we have the results:  the
deceleration-acceleration transition occurs at $z=0.54$.  For the
present universe, $t_0=14Gy=8.20\times 10^{60}M_P^{-1}$, $H_0=74
km/sMpc=1.30\times 10^{-61}M_P$, it is found that \be
    H(t)=\frac{1.066}{t},\ \ \ \ \ \  a(t) \sim t^{1.066}. \label{fb19}
\ee The deceleration parameter $q_0\equiv H_0^{-2}(\ddot
a/a)_0=-0.06191.$ \\
\vp

{\sl The future evolution of the universe} \ \ \ \  After this
transition period we can find  an analytic solution. In fact, when
$t>>10^{60}M_P$, the $\rho_{DE}$ dominates in the righthand
  side of equation(15), then $f(t)\sim t^{-2}$, we have solutions
\be
    H(t)=\frac{4(1-2\gamma)}{3t}=\frac{1.312}{t}, \label{fb20}
\ee\be
    a(t)=a_0 \left(\frac{t}{t_0}\right)^{1.312}  \label{fb21}
\ee and
\be
    f(t)=f_0 \left(\frac{t_0}{t}\right)^2,\ \ \ \ \ \  \rho_m(t)=\rho_{m0 }
    \left(\frac{t_0}{t}\right)^2 . \label{fb22}
\ee \\
The deceleration parameter $q=-(n-1)/n=-0.2378$ and the equation
of motion $w=-0.49$. Therefore
$\Omega_m+\Omega_{DM}=(1+q)/(2-q)=0.34$ \ \ and we have \
$\Omega_{DE}=0.66$, $\Omega_{DM}=0.3$ \ and \ $w_{DE}=-0.75$. \\

\vp In conclusion, we have shown that our universe was dominated
by the energy density $\rho_\phi$ of the gravitational scalar
field with a negative pressure $p_\phi=-\rho_\phi$ in the period
of inflation, afterwards then dominated by a cold dark matter
$\rho_{DM}\simeq\rho_\phi$ after the inflation. After the time of
$10^{60} M_p^{-1}$, the dark energy $\Omega_{DE}$ increased to
accelerate the universe.  Eventually our universe will expand with
a power law $a(t)\sim t^{1.3}$ with $\Omega_{DE}=0.66$ and
$w_{DE}=-0.75$.
 \vp\vp\vp\vp
\begin{center}
{\bf Reference}
\end{center}
\begin{enumerate}
\item
 A.G. Riess {\sl et al.,}  Astrophys. J. {\bf116} (1998) 1009. \ \
 S. Perlmutter et al., Astrophys. J. {\bf517} (1999) 565. \ \ C. Pryke {\sl et
 al.,} Astrophys. J. {\bf 568} (2002) 46. \ \ L. Verde {\sl et al.,} MNRAS {\bf
 335} (2002) 432.  \ \ N. Bahcall {\sl et al.,} Science {\bf 284} (1999) 1481.
 \ \ D.N. Spergel {\sl et al.,} Astrophys. J. Suppl. {\bf 148} (2003)175.
\item
 C. Wetterich, Nucl. Phys. B {\bf 302} (1988) 668. \ \  I. Zlatev {\sl et
 al.,} Phys. Rev. Lett. {\bf 82} (1999) 896.
\item
 G.N. Felder {\sl et al.,} Phys. Rev. D {\bf 66} (2002) 023507.
\item
 M. WANG, Phys. Rev. D {\bf 61} (2000) 123511.
\end{enumerate}
\end{document}